\documentclass[prl,twocolumn,floatfix,nobibnotes,superscriptaddress,prb]{revtex4}
\usepackage{amsmath}
\usepackage{graphicx}

\begin{document}
\title{Phase transformation in steel alloys for  magnetocaloric applications; Fe$_{85-x}$Cr$_{15}$Ni$_{x}$ and Fe$_{85-x}$Cr$_{15}$Mn$_{x}$ as prototypes}
           
\author{P. Souvatzis}
\affiliation{Department of Physics and Astronomy, Division of Materials Theory, Uppsala University,
Box 516, SE-75120, Uppsala, Sweden}
\author{E. K. Delczeg-Czirjak}
\affiliation{Applied Materials Physics, Department of Materials Science and Engineering, Royal Institute of Technology, SE-100 44 Stockholm, Sweden}
\author{L. Vitos}
\affiliation{Department of Physics and Astronomy, Division of Materials Theory, Uppsala University,
Box 516, SE-75120, Uppsala, Sweden}
\affiliation{Applied Materials Physics, Department of Materials Science and Engineering, Royal Institute of Technology, SE-100 44 Stockholm, Sweden}
\affiliation{Research Institute for Solid State Physics and Optics, Budapest H-1525, P.O. Box 49, Hungary}
\author{O. Eriksson}
\affiliation{Department of Physics and Astronomy, Division of Materials Theory, Uppsala University,
Box 516, SE-75120, Uppsala, Sweden}
\date{\today }
\begin{abstract}
We here show by first principles theory that it is possible to achieve  a structural and magnetic phase transition in common steel alloys like  Fe$_{85}$Cr$_{15}$, 
by alloying with Ni or Mn. The predicted phase transition is from the ferromagnetic body centered cubic (bcc) phase to the paramagnetic face centered cubic (fcc) phase.
The relatively high average magnetic moment of $\sim1.4\mu_{B}$/atom predicted at the transition suggests that stainless steel potentially can present a magnetocaloric effect strong enough to make these alloys good candidates for refrigeration applications operating at and around
room temperature.
\end{abstract}

\maketitle

The magneto caloric effect (MCE) has,  since its discovery in 1881 \cite{MCE1}, been subject to a renewed interest from the scientific community during the last decade. 
This second coming has principally  been induced by the discovery of the giant MCE (GMCE) in Gd$_{5}$(Si$_{2}$Ge$_{2}$)\cite{GMCE}, which has been followed 
by the study of several different families of magneto caloric materials such as the lanthanide phases RM$_{2}$ (R = lanthanide, M = Al,Co or Ni), Mn(As$_{1-x}$Sb$_{x}$), MnFe(P$_{1-x}$As$_{x}$),
La(Fe$_{13-x}$Si$_{x}$) and the manganites \cite{geshnider}. The basic idea behind the magneto caloric cooling is to adiabatically transfer  a large change in the entropy of the spin-system to the lattice. One of the most important parameters in searching for new materials with potential in magnetocaloric systems is hence to investigate if, at the ordering temperature, a large change in the magnetization takes place; to be precise one should look for a material with a large $({dM \over dT})$ at constant applied field.

\begin{figure}[tbp]
\begin{center}
\includegraphics*[angle=0,scale=0.3]{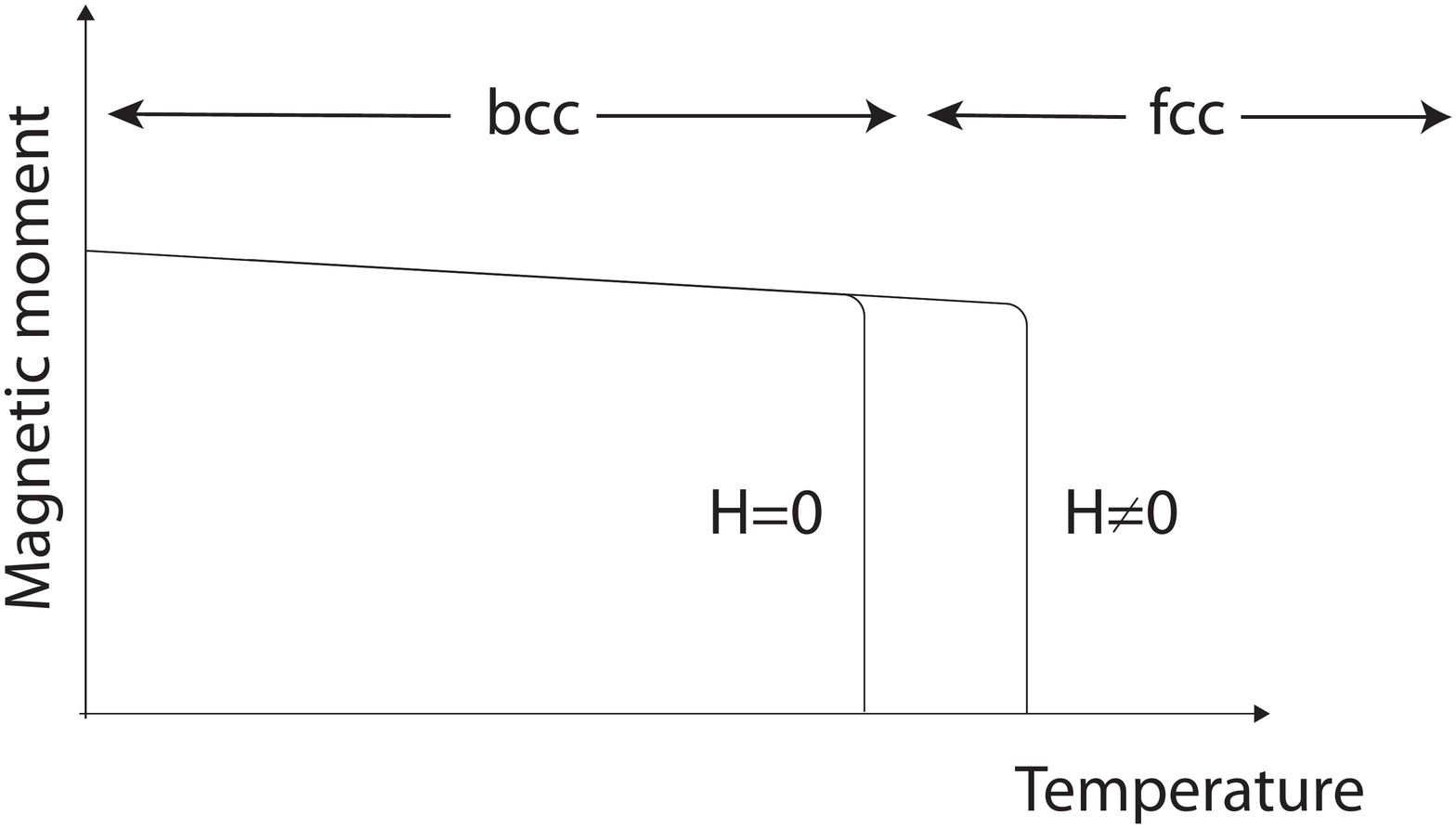}
\caption{Schematic picture of magnetic and structural properties of the discussed materials.}
\label{fig:FIG3}
\end{center}
\end{figure} 

In this paper we propose to add yet another class of materials to the family of compounds with potential to be used in room temperature magnetocaloric refrigeration. The materials suggested herein are the stainless steel alloys Fe$_{1-y-x}$Cr$_{y}$Ni$_{x}$ and 
Fe$_{1-y-x}$Cr$_{y}$Mn$_{x}$ where $0.1<y<0.3$ and $0<x<0.3$. The  idea behind this suggestion is relatively simple and is based upon knowledge that the ferromagnetic bcc-phase  of Fe$_{1-y}$Cr$_{y}$ ( $y\lesssim 20$) can 
be made degenerate with respect to the paramagnetic  fcc-phase  by alloying with Ni or Mn \cite{Ni1,Ni2}. In addition it is known that as a function of increasing temperature, the bcc phase of these alloys becomes destabilized in comparison to the fcc phase \cite{Ni2}, resulting in the ferromagnetic bcc phase being transformed into the paramagnetic fcc phase at high enough temperatures.
Thus, by tuning the chemical composition, it should be possible to tailor these steel alloys so that a  
structural and magnetic transition; from ferromagnetic bcc ($\alpha$ phase) to a paramagnetic fcc ($\gamma$ phase) takes place at any selected temperature in a temperature range around room temperature. This phase transition is expected to be accompanied with a large change in entropy of the spin-system, since the $\alpha$ phase carries a large moment and the $\gamma$ phase is paramagnetic  with smaller disordered local magnetic moment.
A schematic figure of the expected magnetization as a function of temperature in these steels is shown in Fig. \ref{fig:FIG3}, where the bcc $\to$ fcc martensitic transition is schematically shown to be accompanied with a large drop in magnetization, and hence a large value of $({dM \over dT})$. Note that for a material to have good magnetocaloric effect, it should be possible to move this phase with an external field, as shown in the figure.

There is not to our knowledge any previous theoretical nor experimental work focusing on the 
magnetocaloric aspects of these Fe-Cr-Ni and Fe-Cr-Mn alloys, or similar systems, and we have limited our present work  to analyze the likelihood of a new class of 
astonishingly simple magnetocaloric materials, and to hopefully inspire the experimental community to study the Fe$_{1-y-x}$Cr$_{y}$Ni$_{x}$  and Fe$_{1-y-x}$Cr$_{y}$Mn$_{x}$ alloys from a magnetocaloric perspective.

\begin{figure}[tbp]
\begin{center}
\includegraphics*[angle=0,scale=0.33]{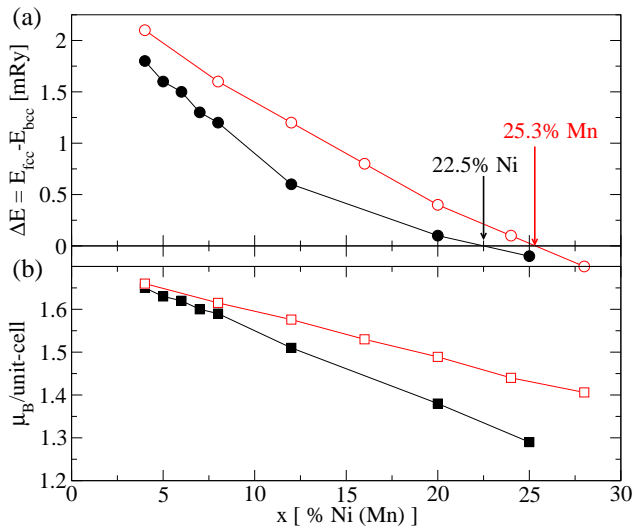}
\caption{(Color online) Calculated total energy differences and magnetic moments obtained within LDA scheme. In (a) the total energy differences  between the bcc and fcc phase of the Fe$_{85-x}$Cr$_{15}$Ni$_{x}$ (filled black circles) and  
Fe$_{85-x}$Cr$_{15}$Mn$_{x}$ (empty red circles) alloys, as function of Ni (Mn) concentration. 
In (b) the calculated magnetic moment of  the Fe$_{85-x}$Cr$_{15}$Ni$_{x}$ (filled  black squares)  and Fe$_{85-x}$Cr$_{15}$Mn$_{x}$ (empty red squares) alloy in the bcc phase, as function of Ni (Mn) concentration.}
\label{fig:FIG1}
\end{center}
\end{figure}

The current work is based on total energy calculations within the frame work of density functional theory, where the main focus is to study the phase stability as well as the magnetic properties of the $\alpha$ and $\gamma$ phase.  
Even though the  Cr concentration has to be carefully chosen to avoid mixing of the $\alpha$ and $\gamma$-phase at temperatures higher than the Curie temperature, we have here for reasons of computational efficiency  limited our study to a Cr concentration of 15\%, merely based on the zero temperature stability of the $\alpha$-phase relative to the $\gamma$-phase in Fe$_{85}$Cr$_{15}$ \cite{Ni1}.

The total energies and magnetic properties of the $\alpha$- and $\gamma$-phase were calculated with the {\it ab initio} exact muffin-tin orbital (EMTO) method \cite{emto1,emto2,emto3,emto4,emto5}.
The electronic correlations were treated within the local density approximation (LDA). Furthermore, all calculations were performed at T=0K neglecting thermal effects. Obviously these are rough approximations, especially for the 
present systems where besides the electron excitations and phonon vibrations, magnetic correlations also play an important role in the phase stability \cite{LDA}. However, we believe that these approximations allow us to provide  
a first qualitative picture on a possible magnetocaloric effect in steels. We also admit that for any quantitative prediction one should go beyond the present LDA and static approach.

In addition to the spin polarized calculations of the $\alpha$- and $\gamma$-phase, paramagnetic calculations  in the $\gamma$-phase was performed within the disordered local magnetic moments (DLM) picture \cite{dlm1,dlm2}. The chemical and magnetic disorder were treated within the coherent potential approximation (CPA) \cite{cpa1,cpa2}. A 13x13x13 Monkhorst-Pack k-point mesh was used.

In Fig.\ref{fig:FIG1} (a) we show the calculated total energy difference between the ferromagnetic $\alpha$ and paramagnetic $\gamma$ phase in Fe$_{85-x}$Cr$_{15}$Ni$_{x}$, and in Fig. \ref{fig:FIG1} (b)
the calculated magnetic moment of the Fe$_{85-x}$Cr$_{15}$Ni$_{x}$ ferromagnetic $\alpha$-phase. Here we see that the two phases become degenerate at  Ni content 
of $\sim$22.5\%. The fcc  paramagnetic phase becoming the more  stable phase near this composition is also in line with the experimental findings of  Majumdar {et al.} \cite{Ni1}. The average atomic magnetic moment of the $\alpha$ phase at the transition is $\sim$1.33$\mu_{B}$, with the individual moments of 1.71$\mu_{B}$ on the Fe-atom, 0.37$\mu_{B}$ on the Cr-atom and 
0.90$\mu_{B}$ on the Ni-atom. In the case of  Fe$_{85-x}$Cr$_{15}$Ni$_{x}$, the DLM-LDA (0K) calculations for the fcc phase predicted an almost vanishing local magnetic moment at all concentration of Ni.

Also in Fig.\ref{fig:FIG1} (a) we show the calculated total energy difference between the ferromagnetic $\alpha$ and the paramagnetic $\gamma$ phase in Fe$_{85-x}$Cr$_{15}$Mn$_{x}$, and in Fig. \ref{fig:FIG1} (b)
the calculated magnetic moment of the Fe$_{85-x}$Cr$_{15}$Mn$_{x}$ ferromagnetic $\alpha$-phase is shown. The the two phases become degenerate at a Mn content 
of $\sim$25\%, and the average atomic magnetic moment of the $\alpha$ phase at the transition is $\sim$1.45$\mu_{B}$. The corresponding individual moments are 1.75$\mu_{B}$ on the Fe-atom, 0.55$\mu_{B}$ on the Cr-atom and 1.26$\mu_{B}$ on the Mn-atom. Also in the case of Fe$_{85-x}$Cr$_{15}$Mn$_{x}$,  the DLM-LDA (0K) calculations for the fcc phase predicted an a almost vanishing local magnetic moment at all concentrations of Mn.
We should emphasize that the predicted transition compositions and magnetic moments have all been obtained within the LDA at 0K. Thus these results should be viewed rater as qualitative than quantitative,
 especially when bearing in mind that it is the magnetic moments around room temperature which is relevant for magneto-caloric refrigeration (see Fig.\ref{fig:FIG3}). 

\begin{figure}[tbp]
\begin{center}
\includegraphics*[angle=0,scale=0.35]{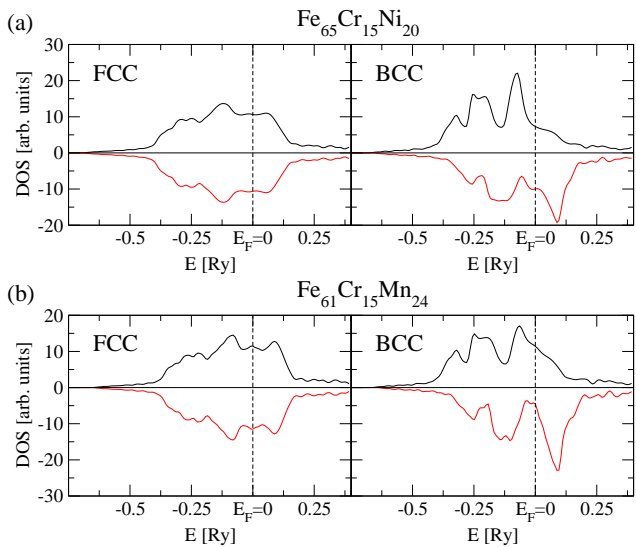}
\caption{(Color online) Calculated density of states obtained within the LDA scheme. In (a) the density of states  of the bcc and fcc phase of the Fe$_{65}$Cr$_{15}$Ni$_{20}$ alloy. 
In (b) the density of states of the bcc and fcc phase of the  Fe$_{61}$Cr$_{15}$Mn$_{24}$ alloy. The density of states for the 
spin down channel is given by the red lines and the spin up by black lines.  }
\label{fig:FIG4}
\end{center}
\end{figure}

The reason for the very different magnetic behaviors of fcc and bcc based Fe systems was discussed already in Ref.\onlinecite{andersen}, and is based on a density of states (DOS) argument. We show the DOS curves of the presently studied systems in Fig.\ref{fig:FIG4}, both for the bcc and fcc phase. In the left column the 
results from the spin polarized calculations for the fcc phase are shown and in the right column  the corresponding results 
for the bcc phase. From the bcc DOS curve it is evident that the non-magnetic bcc phase is unstable relative to ferromagnetic ordering, which is manifested  by  the appearance of an exchange splitting  between the spin-up and the spin-down states, resulting in the peaks of majority spin and the minority spin DOS being pushed down respectively up relative to each other. The fcc phase however, is stable 
relative ferromagnetic ordering, resulting in the spin up and spin down DOS being totally degenerate.

In the present work it has been shown that it is possible to alloy Mn and Ni with Fe-Cr alloys, so that the bcc and fcc phase become degenerate.  It is furthermore shown that the ferromagnetic bcc phase of these alloys has a large magnetic moment and that  the local magnetic moment of the paramagnetic fcc phase  is much smaller than that of the ferromagnetic bcc phase. Hence a substantial magnetic entropy change is to be expected at the event of a bcc to fcc phase transformation.
The near vanishing local moments in the fcc phase, is a result of  the DLM-LDA (0K) calculations. Other approximations, like dynamical mean field theory and generalized gradient approximation, are expected to result in a slightly different scenario, with finite but non-collinear magnetic ordering \cite{LDA}. However, these approximations would also result in a zero net-moment and a M versus T curve in accordance with the schematic picture of Fig. \ref{fig:FIG3}. Nevertheless, more extensive theoretical and/or experimental work are desired in order to establish the magnitude of the entropy change.

From the experimental phase diagram of these alloys it is known that the bcc phase becomes destabilised with elevated temperature. With this knowledge we propose that Fe$_{85-x}$Cr$_{15}$Ni$_{x}$ and Fe$_{85-x}$Cr$_{15}$Mn$_{x}$ alloys might be interesting candidates as magnetocaloric materials. By tuning the Ni (Mn) concentration it should be possible to find an alloy which undergoes a structural and magnetic phase transition at room temperature (or any other temperature where one would like to have a cooling effect), with a large value of  $({dM \over dT})$ at constant applied magnetic field.

Finally, we would like to mention that at small Ni content it is the hcp phase that is the more stable phase and not the fcc phase \cite{HCP1}.
The hcp phase is  nonmagnetic (having zero local magnetic moments) and thus
the magnetic entropy drop would, as going from bcc to hcp,  be even larger than for the present bcc-fcc transition. This observation raises the question for future theoretical and experimental  studies.
Namely, finding a compositional range where also the nonmagnetic hcp and the ferromagnetic  bcc phases are degenerate at ambient pressure.

We acknowledge the Swedish Research Council, the Swedish Energy Agency  and the Foundation for Strategic Research for financial support. Calculations provided by a SNAC allocation. O.E. is also grateful to the ERC (grant  247062 - ASD ) as well as the KAW foundation.
Valuable discussions with Per Nordblad and Matthias Hudl are ackowledged. 

\end{document}